\documentclass{article}
\begin{document}

\begin{center}
\centerline{\large \bf Comment on "Beam-splitters don't have memory:} 
\centerline{\large \bf a comment on "Event-based corpuscular model}
\centerline{\large \bf for quantum optics experiments" by K.Michielsen et al."}
\end{center}

\vspace{3 pt}
\centerline{\sl V.A. Kuz'menko\footnote{Electronic 
address: kuzmenko@triniti.ru}}

\vspace{5 pt}

\centerline{\small \it Troitsk Institute for Innovation and Fusion 
Research,}
\centerline{\small \it Troitsk, Moscow region, 142190, Russian 
Federation.}

\vspace{5 pt}

\begin{abstract}

R. Ionicioiu in arXiv:1012.0647 claims that beam-splitters do not have 
memory. This is the unproved statement. From other side, such small quantum 
objects as molecules, atoms and even photons have memory, which is connected 
with the inequality of forward and reversed processes in quantum physics.
       
\vspace{5 pt}
\end{abstract}

\vspace{12 pt}

The authors of the event-based corpuscular model [1] believe that it is not 
concerned with an interpretation or an extension of quantum theory. However, 
the problem of interpretation of quantum theory immediately appears in the 
R. Ionicioiu's comment [2], in which he proposes experimentally verify the 
question of possible existence of a memory in a beam-splitter. 

We want here to take notice of existence for many years other quite similar, 
but simpler and much more important problem of experimental measurement of 
differential cross-sections of forward and reversed transitions in quantum 
physics. We have sufficient quantity of direct and indirect evidences that 
this values differ in many orders of magnitude (although its integral 
cross-sections should be equal) [3]. This inequality is connected with a 
memory of quantum system about its initial state and this memory manifests 
itself by extremely high cross-section of a reversed transition. 

The excellent experimental evidence of such kind for photons was published 
in [4, 5]. Unfortunately, the authors are not worried by interpretation of 
quantum physics. So, we should make it for them. At the first stage of the 
experiments the narrowband laser radiation pumps the nonlinear crystal, 
which produces two broadband down-converted beams (signal and idler beams). 
So, the high frequency photon is split on the two low frequency photons. At 
the second stage the two broadband down-converted beams are mixed in the 
nonlinear crystal (sum frequency generator) again [4] or induce a two-photon 
excitation process in a rubidium atoms [5].  We can expect that after such 
mixing we shall see broadband high frequency radiation with spectral width 
equal to the sum of the spectral widths of the signal and idler beams. 
However, the discussed experiments [4, 5] show that the same narrowband 
initial frequency radiation appears. And the important additional condition 
should be satisfied: the arrival times of the entangled photons should be 
equal. 

In some sense this experiments are similar to the widely known experiments 
with the detection of a polarized entangled down-converted photons (Alice 
and Bob) [6]. Moreover, the discussed experiments clearly demonstrate an 
important property of the entangled photons: the cross-section of the 
reversed process of their mixing is much greater than the cross-section of 
forward process of their mixing with any other photons. It means that 
down-converted photons have some memory about its initial state. 

Similar situation is with atoms and molecules. The term "coherent 
superposition of internal quantum states" of atoms or molecules exists in 
quantum physics for many years [7]. The experimentalists frequently define 
it as an ability of quantum system to exist in a mixture of different 
quantum states. This is a full nonsense. Quantum transitions are very fast 
and a quantum system in any moment of time exists only in one quantum state. 
However, the quantum system can have some memory about the initial state. 
This memory manifests itself in high cross-section of the reversed process. 
So, the scientists erroneously accept the properties of this memory as a 
manifestation of the properties of mixture of quantum states.

The task of experimental study of differential cross-sections of forward 
and reversed processes in quantum physics is not very difficult. Our 
scientists should only at last wake up and stop to trust in the old myth 
that lows in quantum physics are invariant under time reversal [8, 9].

\end{document}